\documentclass[prb,amsmath,twocolumn,amssymb,floatfix,showpacs,superscriptaddress,footinbib]{revtex4-1}
\pdfoutput=1
\usepackage[dvips]{graphics}
\usepackage{epsfig}
\usepackage{enumerate}
\usepackage{amsmath,amssymb,amsthm}
\usepackage{color}
\usepackage{sidecap}
\usepackage{graphicx}
\usepackage{standalone}
\usepackage{soul}
\usepackage[colorlinks=true,linktoc=page,linkcolor=red,citecolor=blue]{hyperref}

\newcommand{\etal}{{\it et al.~}}
\newcommand{\ie}{{i.e.,~}}

\newcommand\bea{\begin{eqnarray}}
\newcommand\eea{\end{eqnarray}}
\newcommand\beq{\begin{equation}}  
\newcommand\eeq{\end{equation}}

\DeclareMathOperator{\sech}{sech}
\begin{document} 
\title{Signatures of interfacial topological chiral modes via RKKY exchange interaction in Dirac and Weyl systems}

\author{Ganesh C. Paul}
\email{ganeshpaul@iopb.res.in}
\affiliation{Institute of Physics, Sachivalaya Marg, Bhubaneswar-751005, India}
\affiliation{Homi Bhabha National Institute, Training School Complex, Anushakti Nagar, Mumbai 400085, India}

\author{SK Firoz Islam}
\email{firoz.seikh@aalto.fi}
\affiliation{Department of Applied Physics, Aalto University, P.~O.~Box 15100, FI-00076 Aalto, Finland}

\author{Paramita Dutta}
\email{paramita.dutta@physics.uu.se}
\affiliation{Department of Physics and Astronomy, Uppsala  University, Box 516, S-751 20 Uppsala, Sweden}

\author{Arijit Saha}
\email{arijit@iopb.res.in}
\affiliation{Institute of Physics, Sachivalaya Marg, Bhubaneswar-751005, India}
\affiliation{Homi Bhabha National Institute, Training School Complex, Anushakti Nagar, Mumbai 400085, India}

\begin{abstract}
We theoretically investigate the features of Ruderman-Kittel-Kasuya-Yosida (RKKY) exchange interaction between two magnetic impurities, mediated by the interfacial bound states inside a domain wall (DW). The latter separates the two regions with oppositely signed inversion symmetry broken terms in graphene and Weyl semimetal. The DW is modelled by a smooth quantum well which hosts a number of discrete bound states including a pair of gapless, metallic zero-energy modes with opposite chiralities. We find clear signatures of these interfacial chiral bound states in spin response (RKKY exchange interaction) which is robust to the deformation of the quantum well.
\end{abstract}

\maketitle
\section{Introduction}{\label{sec:I}}
The emergence of the gapless one dimensional ($1$D) chiral modes across the interface of two non-equivalent  trivial or topological insulators has received significant attention owing to its potential  application as one way wave propagation in two-dimensional ($2$D) honeycomb photonic lattice\,\cite{haldanePRL,haldanePRA} as well as spin and valley selective charge transport in $2$D Dirac materials\,\cite{DW_silicene,Ezawa_silicene}. The underline physics lies in the sign change of the symmetry breaking parameter across the interface, forming a quantum well (QW) which acts as a domain wall (DW) separating the two insulators. The chiralities of these interfacial modes, corresponding to the two valleys of the Dirac materials, are sensitive to the types of broken symmetry. The time reversal symmetry (TRS) breaking leads to the same chirality of the interfacial modes\,\cite{haldanePRL}, whereas they appear with opposite chirality for inversion symmetry (IS) broken systems\,\cite{niu,semenoff-gapped,cayssol}. Such interfacial chiral modes (ICM) have also been found in bilayer graphene\,\cite{blg_Martin,Blg_McDonald1,Blg_McDonald2,blg_Wang} which motivated to propose a bilayer graphene based Cooper pair beam splitter with maximum efficiency\,\cite{Recher}. Subsequently, similar investigations were carried out in silicene\,\cite{DW_silicene}, surface of three-dimensional ($3$D) topological insulators\,\cite{3dTI_Torres}. Very recently, a pair of ICM have been revealed in $3$D spin-1 topological semimetal where TRS is broken by means of light\,\cite{islam3D}.

In recent times, the intriguing topological properties of 3D Weyl semimetal (WSM) have attracted a great deal of attention in the research community \cite{Weyl_ViswanathSavrasov,Weyl_BurkovBalent,ZyuzinBurkov,ViswanathRMP,TaAs,chen2019optical1,chen2019optical1}. One major focus is the IS broken WSM where the valence and conduction bands touch each other at minimum four 
or more topologically protected Weyl nodes\,\cite{Weyl_ViswanathSavrasov,ViswanathRMP,TaAs}. These nodes are separated both in momentum and energy and the low energy spectrum which is linear around these nodes, is described by the Weyl equations. Each Weyl node of the WSM manifests definite chirality with a total of zero as it always appears in pairs in the momentum space\,\cite{Nielson}. 
The chirality property of the bulk band and the surface Fermi arcs lead to several exotic phenomena in WSM\,\cite{Parameswaran,CA-Burkov,CA_Carbotte,CA-Hasan,Stern_PRX,Zyuzin-SC,PD_odd-Weyl,chen2019optical2}.

Along this framework, in this article, we analytically show the emergence of ICM in IS broken WSM. 
\begin{figure}[htb]
\centering
\includegraphics[scale=0.37]{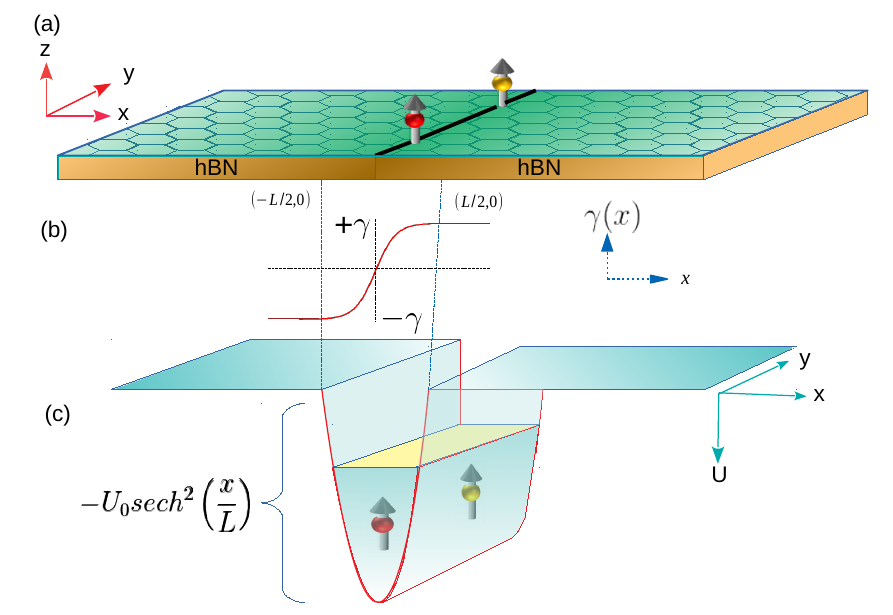}
\caption{(Color online) (a) A schematic sketch of graphene, placed on the top of a hBN substrate, is depicted. The two magnetic impurities are placed on the DW which is denoted by the black solid line. 
(b) The smooth variation of the mass term, modelled by $\gamma(x)=\gamma\tanh(x/L)$, is shown. (c) QW of width $L$ and height $U_0$, defined by the hyperbolic function, 
developed across the DW is symbolically illustrated.}
\label{model}
\end{figure}
Probing these ICM remains always challenging because of its geometric confinement within the DW\,\cite{kang2018pseudo}. In the present work, we also aim to extract the direct signatures of these ICM via Ruderman-Kittel-Kasuya-Yosida (RKKY)\,\cite{ruderman1954indirect,kasuya1956electrical,yosida1957magnetic} exchange interaction between two magnetic 
impurities placed across the DW created in gapped graphene and in WSM with broken IS. It is an indirect exchange interaction mediated by the conduction electrons of the host material and already investigated extensively in different Dirac materials\,\cite{grapheneABS,kogan,gorman,satpathy1,satpathy2,xiao2014ruderman,Dassharma}, topological insulator\,\cite{zhusurfaceTI} etc.   
RKKY exchange interaction has also been proposed to determine the magnetic ordering in spin glasses\,\cite{spinglass}, alloys\,\cite{alloys} and to probe topological phase in silicene\,\cite{rkky_bulkSilicene}, edge states of graphene nanoribbon\, \cite{PhysRevB.87.045422} and 2D topological insulator\, \cite{PhysRevB.96.081405}, decoupled edge modes in phosphorene\,\cite{FPA-rkky}, order of tilting in spectrum of borophene\,\cite{GFA-rkky} and Fermi arc in WSM thin films\,\cite{PhysRevB.101.085419} etc. 
Several experimental methods like single-atomic magnetometry and magnetotransport measurement based on angle-resolved photo-emission spectroscopy (ARPES) are used to capture this 
exchange interaction\,\cite{PhysRevLett.91.116601,zhou2010strength,khajetoorians2012atom}.

The remainder of this paper in organized as follows. In Sec.~\ref{sec:II}, we shortly revisit the appearence of ICM in graphene due to IS breaking mass term accross the interface.
In Sec.~\ref{sec:III}, we discuss the emergence of ICM due to oppositely signed momentum shifts in inversely broken WSM. The stability of both massless and massive interfacial modes against the deformation of QW is shown in Sec.~\ref{sec:IV}. Sec.~\ref{sec:V} is devoted to the discussion of signatures of these ICMs via RKKY exchange interaction (spin response). Finally, we summarize and 
conclude in Sec.~\ref{sec:VI}.

\section{Model and ICM in $2$D graphene}{\label{sec:II}}
We begin by revisiting a gapped graphene Hamiltonian\,\cite{semenoff-gapped} $H_{2\text{D}}=\sigma\cdot k+\gamma (x)\sigma_z$, with $\sigma\equiv\{\sigma_x,\sigma_y\}$ where $\sigma_i$'s ($i\in \{x,y,z\}$) are Pauli 
matrices in the sublattice space and $k\equiv\{k_x,k_y\}$ is the $2$D momentum operator. The mass term $\gamma(x)$ is responsible for breaking IS, which can be practically realized by placing the monolayer graphene on 
the top of a hexagonal boron-nitride (hBN) substrate\,\cite{bandgap-hBN,novoselov} as depicted in Fig.\,\ref{model}
(a). We model the mass term as $\gamma(x)=\gamma \tanh(x/L)$ so that it smoothly changes its sign across a 
region, namely, DW of width $L$ around $x=0$ and show in Fig.\,\ref{model}(b). The DW separating the two 
regions of oppositely signed mass terms is marked by black solid line in Fig.\,\ref{model}(a). Consideration of the 
smoothly varying mass term is well justified as it is a formidable task to design sharp boundary in reality. After 
decoupling the square of the eigenvalue equation, $H_{2\text{D}}\Psi=E\Psi$, we obtain
 \begin{equation}
  \left[-\frac{\partial^2}{\partial (x/L)^2}+U\left(\frac{x}{L}\right)\right]\psi_i=L^2(E^2-k_y^2-\gamma^2)\psi_i\ ,
  \label{Eq:1}
\end{equation}
where $i \in$ \{A, B\} is the sublattice index. It resembles an $1$D Schr\"{o}dinger equation for well-known P\"{o}schl-Teller QW of width $L$ as\,\cite{lekner2007,diaz}
 \begin{eqnarray}
U\left(\frac{x}{L}\right)=-\gamma L(\gamma L+1) \sech^{2}\left(\frac{x}{L}\right)\ ,
\label{eq:PT}
 \end{eqnarray}
and schematically shown in Fig.\,\ref{model}(c). The QW effectively describes the DW with the bound state energy given as\,\cite{landau2013quantum}
\begin{subequations}
\begin{eqnarray}
E_{n=0}&=&sgn(\gamma)k_y\\
E_{n>0}&=&\pm \sqrt{k_y^2+2|n| \frac{\gamma}{ L}-\frac{n^2}{L^2}}~.
\end{eqnarray}\label{2dband}
\end{subequations}
Note that, $sgn(\gamma)$ denotes the chirality of the {\it {zeroth interfacial bound states}} and the massive mode solutions are valid for $2\gamma L> |n|$. 
The normalized wave function is given by\,\cite{landau2013quantum,freitas} $\psi_{n,k_y}({\bf r})=[e^{ik_yy}/\sqrt{L_y}]\phi_n(x/L)$ with
\begin{equation}
 \phi_{n}\left(x\right)=\frac{A_n}{\left[\cosh(x)\right]^{b}}P_n^{b,b}\left[\tanh \left(x\right)\right]\ ,
 \label{psi_2d}
\end{equation}
where $P_n^{b,b}(x)$ is the Jacobi polynomial with $b=\gamma L-n$. The normalization factor is given by
\begin{equation}
A_n=\sqrt{\frac{n!b}{2^{2b}}\frac{(2b+|n|)!}{(\gamma L!)^2}}\ ,
\label{eq:An}
\end{equation}
with $n_{max}<\gamma L$.
Note that, the chiralities of the {\it {zeroth interfacial bound states}} \ie $E_{n=0}$ (Eq.(\ref{2dband}(a))) at the two valleys are opposite to each other. Even though IS symmetry breaking does not lead to the topological phase transition, the opposite chirality is still topologically preserved as the difference of left and right moving chiral modes must be equal to the difference of the Chern numbers between left and right regions separated by the DW\,\cite{haldanePRA,cayssol}. Alternatively, one can break the TRS in the two regions of the graphene where only unidirectional ICM emerges at the two valleys\,\cite{haldanePRL,haldanePRA,cayssol} and thus RKKY exchange can arise nonlocally along the other edge or surface in a finite system\,\cite{Kundu}. In our case, these gapless ICM appear due to the sign change of the mass term across the boundary. They are topologically protected by the TRS present in the system and thus insensitive to how smooth sign change is (even valid for 
abrupt change) and also other details of the QW.  However, this topological protection is absent for $n>0$ modes being sensitive to the dimension ($L$) and the smoothness of the QW as clearly seen from (Eq.\,\ref{2dband}(b)). 
We note that these $n>0$ modes are similar to the Volkov and Pankratov states\cite{volkov1985two} that arise at the interface of two semiconductors/band insulators with inverted band gaps 
(mass terms)\,\cite{tchoumakov2017volkov,mukherjee2019dynamical,van2020volkov}.

\section{ICM in $3$D Weyl Semimetal}{\label{sec:III}}
Here, we explore the appearance of ICM in IS broken $3$D WSM. The IS broken WSM hosts four inequivalent Weyl nodes in the momentum space.  The interfacial modes in 3D topological hetererojunction have been considered previously with inverted mass term\cite{tchoumakov2017volkov,mukherjee2019dynamical} which changes sign across the interface and opens 
mass gaps on the both sides of the interface \ie a junction is made between two insulating phases. The ICM for electromagnetic waves has also been predicted in a TRS broken WSM\,\cite{PhysRevB.92.115310}. However, in this work we consider an interface between two IS broken WSMs with oppositely signed momentum shifts instead of mass. Note that, the momentum shift does not open any gap unlike the previous studies\,\cite{tchoumakov2017volkov,mukherjee2019dynamical}, and hence the topological character of the Weyl nodes still persist. The low energy effective Hamiltonian for one of these Weyl nodes reads as\,\cite{Trauzettel}
\begin{equation}
 H_{3\text{D}}=k_x \tau_x+(k_y-k_0) \tau_y+[k_z-\beta(x)] \tau_z\ ,
\end{equation}
where $k_i$'s are the momentum operators, $k_0$ is the system parameter and $\tau_i$'s are the Pauli matrices acting 
on the spin basis with $i\in\{x,y,z\}$. Here, $\beta(x)$ is the IS breaking term and unlike graphene, it does not open any gap but shifts the Weyl nodes along $k_z$ direction which is 
evident from the energy dispersion as $\epsilon_k=\sqrt{k_x^2+(k_y-k_0)^2+(k_z-\beta)^2}$. The gapless Weyl nodes are topologically protected even after the 
IS or TRS symmetry breaking perturbation, provided translational symmetry is preserved. Following the same prescription as in graphene, we consider the IS breaking term as $\beta(x)=\beta\tanh(x/L)$ which smoothly changes its sign across the interface. After squaring the eigenvalue equation, $H_{3\text{D}}\Psi=E\Psi$, we arrive at
\begin{equation}
 \left[-\frac{\partial^2}{\partial x^2}+(\beta(x)-k_z)^2-\tau_y\frac{\partial \beta(x)}{\partial x}\right]\Psi=[E^2-(k_y-k_0)^2]\Psi\ .
\end{equation}
It can be decoupled as
\begin{equation}
 \left[-\frac{\partial ^2}{\partial (x/L)^2}+V \left(\frac{x}{L}\right)\right]\psi_j=L^2[E^2-\beta^2-k_z^2-(k_y-k_0)^2]\psi_j\ ,
 \label{eq:SE-3d}
\end{equation}
where $j$ denotes spin index and
\begin{equation}
 V\left(\frac{x}{L}\right)=-\beta L\left(\beta L+1\right)\sech^2\left(\frac{x}{L}\right)+2 k_z \beta L^2\tanh\left(\frac{x}{L}\right) \ .
\end{equation}
This is the well-known Rosen-Morse potential\,\cite{Morse1,Morse2} which exhibits a minimum for $|2k_z L|<(\beta L+1)$, indicating a region of $k_z$ favouring the QW rather than potential barrier. 
Note that, the first term is exactly similar to Eq.\,(\ref{eq:PT}) with $\beta$ replaced by $\gamma$. By drawing the analogy with Rosen-Morse's QW solution, we obtain the bound state solutions with 
the energy eigenvalues as
\begin{subequations}
\bea
 E_{n=0}&=&sgn(\beta)(k_y-k_0)\ , \\
 E_{n>0}&=&\pm\sqrt{\left(\frac{2n\beta}{L}-\frac{n^2}{L^2}\right)\left[1-\left(\frac{k_zL}{\beta L -n}\right)^2\right]+(k_y-k_0)^2}\ , \nonumber\\
\eea  \label{band_3d}
\end{subequations}
where, the {\it zeroth} solution yields the ICM ($E_{n=0}$) and the other solutions represent the massive modes ($E_{n>0}$) in WSM. This is one of the main results of our paper. The corresponding eigenstates are given by $\psi_{n,k_y,k_z}({\bf r})=[e^{i(k_yy+k_z z)}/\sqrt{L_y L_z}]\tilde{\phi}_n(x/L)$ with
\begin{equation}
 \tilde{\phi}_n(x)=A_n\frac{e^{-a x}}{[\cosh(x)]^b}P_n^{b-a,b+a}[\tanh(x)]\ ,
 \label{wf_3d}
\end{equation}
where, $b\pm a=\sqrt{(\beta\pm k_z)^2+(k_y-k_0)^2-E_{n}^2}$. This reduces to $b=\beta$ and $a=k_z$ for {\it {zeroth}} ICM. The normalization factor is given as follows
\begin{equation}
A_n=\sqrt{\frac{n!}{2^{2b}}\frac{(2b+n)!}{[(b+a+n)!(b-a+n)!}\frac{b^2-a^2}{b}}\ ,
\label{eq.:Norm3D}
\end{equation}
with $n_{max}<\beta L-L\sqrt{\beta k_z}$. Interestingly, the above band structure boils down to that of graphene (see Eq.\,(\ref{2dband})) or in other words, the bound state solutions for Rosen-Morse QW reduce to the P\"{o}schl-Teller QW solutions for $k_z=0$. Note that, the zeroth ICM are immune to the details of interface and it can be obtained directly from the abrupt interface, discussed later. One can also find solutions for the other three Weyl nodes in a similar way. 

\section{Stability of ICM}{\label{sec:IV}}
Now, we discuss the stability of the ICM against the deformation and abruptness of the QW.
So far, our results are based on smooth QW for both graphene and WSM.  In reality, the QW may not be ideally described by the hyperbolic function because of the weak deformation which might arise from the asymmetry in the smoothness of the IS breaking term on both sides of the interface. Such weak deformation can be taken into account by considering $q$-deformed hyperbolic function, introduced by Arai\,\cite{ARAI199163} as $\sinh_q(x)=(e^x-qe^{-x})/2$ and $\cosh_q(x)=(e^x+qe^{-x})/2$ with $0<q<1$. This follows $q \sech^2_q(x)+\tanh_q^2(x)=1$, $\cosh_q^2(x)-\sinh_q^2(x)=q$ and $\frac{d}{dx} \tanh_q(x)=q \sech_q^2(x)$. Hence, we model our inversion breaking term by $q$-deformed hyperbolic function as $\gamma(x)=\gamma \tanh_q(x/L)$ which yields $q$-deformed Rosen-Morse potential 
given as
\beq
V_q(x)=-q\beta L\left(\beta L+1\right) \sech_q^2(x)+2 k_z\beta L^2\tanh_q(x)\ .
\eeq
Interestingly, noting the analogy with the solutions of $q$-deformed Rosen-Morse potential\,\cite{eugrifes}, we find that the energy spectrum does not depend on $q$. Hence, we confirm that not only the massless ICM even the massive mdoes are robust to such deformation of the QW. This robustness of the ICM elevates the potential for application and thus highly enhances the importance of the present work. Similar to the Rosen-Morse QW for WSM, the deformed P\"{o}schl-Teller QW corresponding to the gapped graphene also yields bound state solutions that do not depend on the degree of deformation\,\cite{deformed_soln}. Therefore, ICM in both graphene and WSM are robust to such deformation of the QW.

Furthermore, we show that the {\it zeroth} ($E_{n=0}$) ICM can be even directly obtained by considering an abrupt interface. We model the abrupt interface as $\beta(x)=-\beta[2\Theta(x)-1]$, 
where $\Theta(x)$ is a Heaviside step funcion. The eigenvalue equation can be decoupled to write
\begin{equation}
 \left[-\frac{\partial^2}{\partial x^2}-2\beta\delta(x)\right]\psi_j=[(k_y-k_0)^2+(k_z-\beta(x))^2-E^2]\psi_j\ ,
\end{equation}
which is an $1$D delta-function QW problem. By employing the appropriate boundary condition across the interface at $x=0$, we obtain an equation to determine the energy dispersion as
\begin{equation}
 \sum_{\eta=\pm}\sqrt{(k_y-k_0)^2+(k_z+\eta\beta)^2-E^2}=2\beta\ ,
\end{equation}
satisfying the interfacial bound states $E=sgn(\beta) (k_y-k_0)$ which is exactly same as Eq.\,(\ref{band_3d})(a).

The wave functions corresponding to the ICM is $\Psi_{+}\propto \exp(-x/l_{+})$ for $x>0$ and $\Psi_{-}\propto\exp(x/l_{-})$ for $x<0$, where $l_{\pm}=\lvert k_z\pm\beta\rvert^{-1}$ 
are the localization lengths of ICM, which is not symmetric with respect to the interface. Here we quickly comment that in graphene, the localization length is $|\gamma|^{-1}$ and thus symmetric 
about the interface.

Note that, the IS breaking WSM Hamiltonian is considered in real spin space, hence it is interesting to examine how the spin configuration responds to the sign changes of the momentum shift across the interface. Using the eigen states for the ICM across the interface, we find the spin polarization for three components as: $\Psi^{\dagger}\tau_{x}\Psi=0$, $\Psi^{\dagger}\tau_{y}\Psi=sgn[E_0+(k_z-\beta sgn(x))]$ and $\Psi^{\dagger}\tau_z\Psi\propto [1-sgn(x)]$, which indicate that the spin polarization across the interface exhibits a sharp discontinuity along the direction of the momentum shift. For graphene, we can immediately conclude that the pseudo-spin polarization possesses similar discontinuity.
\section{Signatures of ICM in RKKY exchange interaction}{\label{sec:V}
To investigate the signature of ICM via a response function in graphene and WSM, here 
we evaluate the RKKY exchange interaction between two magnetic impurities, mediated by the bound states of 
the QW. Numerous works on the RKKY exchange interaction have been carried out so far concering Dirac materials\,\cite{xiao2014ruderman,kogan,grapheneABS,gorman,satpathy1,satpathy2,Dassharma,rkky_bulkSilicene,
rkky_diracWeyl,wang2020rkky}, where exchange interaction takes place via bulk conduction electrons and bound 
states, formed inside a circular QW in graphene~\cite{Canbolat}. It is noteworthy to mention here that the RKKY 
exchange interaction via the bound states of a QW formed in metallic heterostructure was considered by several 
groups in early 90's\,\cite{QW_hanna,QW_barnas}. However, the study is still lacking as far as the signature of 
ICM in graphene and WSM are concerned.  

For our computation, we employ the formalism (based on first quantized language) already developed in Refs.\,\cite{QW_hanna, QW_barnas} for QWs, and very recently used in case of 
graphene\,\cite{Canbolat}. The total Hamiltonian in presence of two magnetic impurities can be written as
\begin{equation}
\mathcal{H}=H_{2\text{D}(3\text{D})}+\Lambda \sum_{i}\mathbf{s}({\bf r}_i). \mathbf{S}_i\ ,
\label{H_lambda}
\end{equation}
where $\mathbf{s}({\bf r}_i)$ is the spin of the conduction electrons at position ${\bf r}_i$ and $\mathbf{S}_i$ is the 
spin of the magnetic impurity. The first part of the Hamiltonian represents the Hamiltonian of $2$D or $3$D system 
under consideration and the second term corresponds to the direct exchange interaction of the conduction electrons, 
of strength $\Lambda$, with the magnetic impurities. In the perturbative limit of $\Lambda$, the second order
contribution to the ground-state energy can be obtained as\,\cite{QW_barnas}
 \begin{equation}
 \Delta E=-\left(\frac{\Lambda}{n_c}\right)^{2} \sum_{i,j}J_{\rm ex}({\bf r}_i,{\bf r}_j)\mathbf {S_i}\cdot \mathbf{ S_j}\ ,
 \label{e_rkky}
 \end{equation}
where, $(\Lambda/n_c)^2 J_{\text{ex}}({\bf r_i,r_j})$ is the RKKY exchange interaction between the impurity spins $i$ and $j$ located at positions ${\bf r}_i$ and ${\bf r}_j$ respectively, and $n_c$ is the conduction electron density. The RKKY interaction strength can be expressed in terms of the wave functions of the unperturbed Hamiltonian as\,\cite{QW_hanna}
\begin{equation}
 J_{\text{ex}}({\bf r_i,r_j})=\frac{1}{2}\sum_{\substack{E_{\xi}>\mu \\ E_{\xi'}<\mu}}\frac{\psi^{\ast}_{\xi'}({\bf r_i})\psi_{\xi}({\bf r_i})\psi^{\ast}_{\xi}({\bf r_j})\psi_{\xi'}({\bf r_j})+\text{h.c.}}{E_{\xi}-E_{\xi'}}\ ,
 \label{f_rkky}
\end{equation}
where, in the wave function $\xi={n,k_y}$ (${n,k_y,k_z}$) for $2$D ($3$D) and $\mu$ is the chemical potential. Although ideally one should take summation over all the energy states inside as well as outside of the QW, we restrict ourselves only to the states inside the QW as we aim to extract the signature of gapless ICMs' close to the undoped situation.
\begin{figure}[htb]
\includegraphics[height=4.5cm,width=9.0cm]{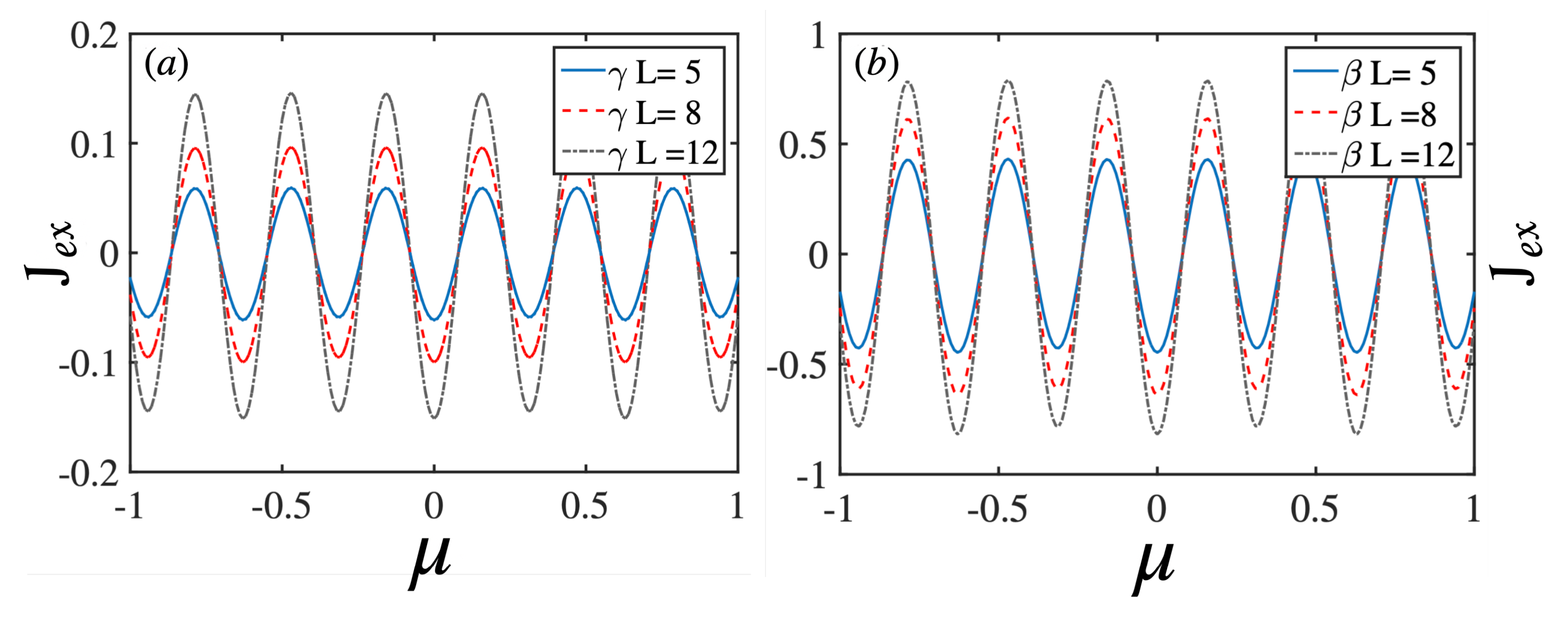}
\vskip-0.3cm
\caption{(Color online) $J_{\text{ex}}$ is depicted as a function of $\mu$ for (a) graphene and (b) WSM with $x=0$ 
and $y=10L$. Here, $\mu$ and $J_{\rm ex}$ are normalized by $\gamma$ and $\gamma L^{2}$ (graphene) or 
$\beta L^2$ (WSM), respectively. }
\label{Fig2:rkky_mu}
\end{figure}

Note that our calculation is valid in the perturbative limit of $\Lambda$ (see Eq.\,(\ref{e_rkky})) where 
the second order contribution to the ground-state energy contains the information of RKKY exchange interaction 
between the impurity spins which break the translational symmetry only locally. Hence, only electrons within a limited wavelength range near the 
Fermi energy are scattered resulting in a density modulation around the impurity. Therefore, the band structure of graphene 
as well as WSM will not be significantly modified by such density oscillations in presence of the impurities. Also, they can't influence the 
topological character of the bulk as well as ICM in 2D graphene and 3D WSM.

In Fig.\,\ref{Fig2:rkky_mu}(a) we show that the RKKY exchange interaction strength for 2D graphene as a function 
of the chemical potential $\mu$ for three different values of the parameter $\gamma L$. We restrict $\mu$ to be 
within the energy window where only gapless ICM appear. We find that the exchange interaction exhibits oscillatory 
nature which is immune to the values of $\gamma L$. This oscillatory nature is quite usual where RKKY interaction 
is mediated by the bulk states\,\cite{satpathy1,grapheneABS}. Though the {\it zeroth} energy ($E_{n=0}$) contribution 
does not depend on $\gamma L$, the exchange interaction smoothly enhances via the probability wave function with 
the enhancement of the IS breaking term. Interestingly, unlike other systems, the behavior of $J_{\text{ex}}$ with 
$\mu$ is non-decaying due to the linearity of the ICM. On the other hand, we have checked that contributions arising 
due to the massive modes ($E_{n>0}$) are 
significantly smaller than that of ICM ($\sim 10^{-3}$). 
Therefore, the RKKY exchange interaction is mainly dominated by the gapless modes, indicating direct signatures of ICM.

The contribution to the exchange interaction $J_{ex}$ is mainly dominated by the ICM for which 
an analytical approximate form can be obtained as $J_{ex}\propto Ci[(k_c-\mu)y]$ where $k_c$ is the momentum 
cut-off and $Ci(x)$ is the cosine integral which, for long distances, can be approximated as
 \begin{equation}
  Ci(\zeta)\simeq\frac{1}{\zeta}\left[\sin(\zeta)-\frac{\cos(\zeta)}{\zeta}\right].
 \end{equation}
Therefore, the long distance behavior is exclusively related to the linear dispersion of the ICM which resembles to the RKKY exchange interaction behavior along the edge modes in topological spin Hall liquid\,\cite{hosseini2020}. On the other hand, along the direction of QW width, the long distance limit is not desired as it is restricted by the QW width. However,
a short distance limit might be further simplified by setting $\tanh(x)\simeq x$ and $\cosh(x)\simeq 1$, which yields $J_{ex}\propto [P_n^{b,b}(x)]^4$ for graphene and 
$J_{ex}\propto e^{-4ax}[P_{n}^{b-a,b+a}(x)]^4$ for WSM, indicating the a relatively faster decay in the later case.
\begin{figure}
\includegraphics[height=4.5cm,width=9cm]{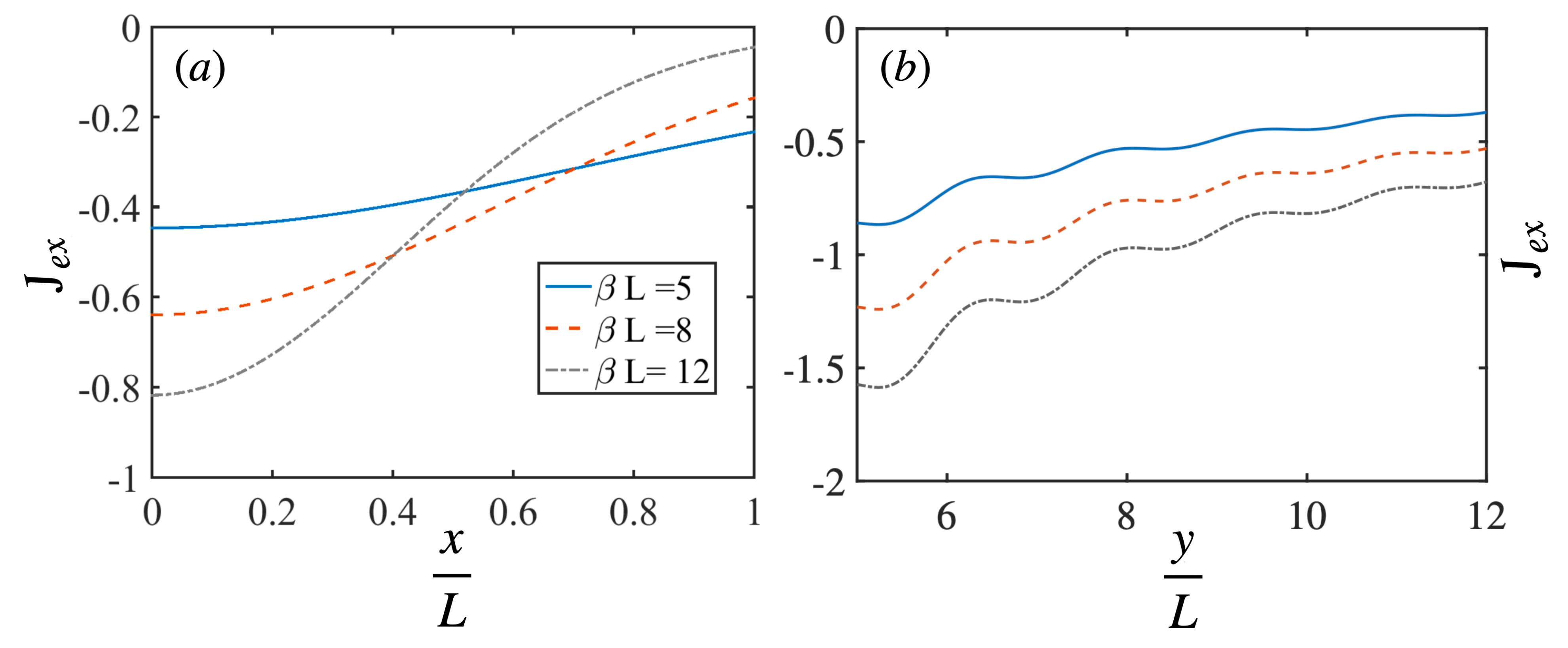}
\vskip -0.2cm
\caption{(Color online) $J_{\text{ex}}$ is demonstrated, in case of WSM, as a function of (a) $x/L$ and (b) $y/L$ for 
various values of $\beta L$ at $\mu=0$. The normalization of $J_{\text{ex}}$ is same as in Fig.\,\ref{Fig2:rkky_mu}.}
\label{Fig3:rkky_xy}
\end{figure}

To further reinforce our results for the signatures of the ICM in RKKY interaction, we extend our study for the $3$D 
system \ie IS broken WSM and present in Fig.\,\ref{Fig2:rkky_mu}(b). The behavior of the RKKY exchange interaction 
as a function of $\mu$ is oscillatory in WSM and the profile is very similar to that of graphene but with an enhanced amplitude. This enhancement of $J_{\text{ex}}$ can be attributed to the prefactor of Eq.\,(\ref{eq.:Norm3D}) appeared 
due to the third dimension of the WSM irrespective of the value of $\beta L$. Unlike graphene, the width of the QW 
has to be maintained quite large so that the bulk states which are gapless in either sides of the QW, do not penetrate 
into the well. In WSM it has been always challenging task to separate the surface modes from the bulk states as both 
are gapless but in our case, a segment of the surface states can only pass through the QW where bulk states are not allowed which might be useful in probing surface modes in WSM. Note that, here the {\it zeroth} ICM are a segment 
of the entire surface states which enclose the WSM.

Finally, we investigate the behavior of $J_{\rm ex}$  for WSM as a function of $x$, the distance from the center of the 
DW and $y$, the separation between the two impurities. We present our corresponding results in Fig.\,\ref{Fig3:rkky_xy}. In Fig.\,\ref{Fig3:rkky_xy}(a), we observe that the contributions of the bound states to the exchange interaction amplitude is maximum at $x=0$ and decays exponentially with $x$ following the hyperbolic form of the potential of the QW. This 
can also be understood following the decay of the probability of the wave functions. This phenomenon is true for all 
$\beta L$. However, the rate of decay is much faster for higher $\beta L$. Here, we set $\mu=0$ to consider only the $E_{n=0}$ ICM. As we increase $\beta L$, the exchange interaction increases following the envelope of $|\psi_{n=0}|^2$ and thus, it further confirms the contributions arising solely due to ICM within the DW. In case of graphene, the RKKY interaction with `$x$' and `$y$' behaves with very similar fashion, except a small mismatch in decay rate which can be easily understood from the absence of $e^{-ax}$ term in graphene ICM wave function.

Furthermore, when we increase the distance between the two impurities placed along the $y$ axis, the magnitude of $J_{\text{ex}}$ decreases with an oscillatory envelope as shown in 
Fig.\,\ref{Fig3:rkky_xy}(b). This behavior of $J_{\rm ex}$ with the distance between the two impurities is very similar to the results for RKKY interaction via bulk states of any doped electronic system\,\cite{satpathy2,Dassharma}. Also, with the enhancement of the IS breaking term, $\beta L$, there is a rise in the strength of RKKY interaction following the envelope of $|\psi_{n=0}|^2$. Note that, along both $x$ and $y$ directions, $J_{\rm ex}$ exhibits anti-ferromagnetic RKKY exchange ($J_{\rm ex}<0$) at $\mu=0$ as the $E_{n=0}$ ICM carries two propagating modes with opposite chiralities resulting in back-scattering with anti-ferromagnetic character.

\section{Summary and Conclusions}{\label{sec:VI}
To summarize, in this article, we have identified the emergence of ICM in IS broken graphene (2D) and WSM (3D). 
We also explore the chiral bound states mediated RKKY exchange interaction between two magnetic impurities 
placed in them. The two magnetic impurities are localised inside a DW separating the two regions with oppositely 
signed IS breaking semenov mass term in graphene and momentum shifting in WSM. We model the DW by a 
hyperbolic potential well. Most interestingly, in the undoped case ($\mu=0$), we have found clear signatures of the 
ICM ($E_{n=0}$) of the gapped graphene and WSM in the RKKY exchange interaction due to their dominating contributions compared to the other bound states ($E_{n \neq 0}$) of the QW localised within the interfacial region. For any condensed matter system, it is always challenging task to isolate surface states from the bulk states. In our 
case, the unique ICM are completely separated from the bulk due to the finite size of the QW engineered in both, graphene and WSM.

In recent times, a few proposals have been put forwarded to extract the contributions of the Fermi arcs 
 which appear across the interface between the WSM and vacuum via RKKY exchange interaction\,\cite{PhysRevB.101.085419,duan2018indirect}. The contributions of the bulk states are 
 inevitably present in those studies when the impurities are located at two opposite surfaces. On the contrary, in our case, the RKKY exchange interaction is separated from the bulk states, 
 bearing the contributions of only the ICM confined inside the DW.

From the practical point of view, TRS and IS broken WSM can be realized in $\rm TaAs, TaP$\,\cite{ScienceWSMExp1,ScienceWSMExp2}. Magnetic adatoms (Fe, Co etc.) can be implanted in materials
to induce magnetic moments in them\,\cite{PhysRevLett.101.137206}. 
An atomically precise map of the magnetic coupling between individual adatoms in pairs can be revealed in terms of spin-resolved differential conductivity following the investigation by  Zhou \etal on the RKKY interaction by depositing Co adatoms on Pt ($111$) surface\,\cite{zhou2010strength}. They extracted the out-of-plane components of the time-averaged magnetization, both in absence and presence of external magnetic field, to find the interaction strength. For our model, the RKKY exchange interaction strength $J_{ex}$ can also be measured in the similar way after successfully depositing adatoms on graphene and/or WSM. The distance dependent behavior of $J_{ex}$ can be found by maintaining different distances between the two impurities. Spin-dependent scanning tunnelling spectroscopy\,\cite{ScienceRKKYExp,zhou2010strength} can be used to measure $dI/dV$ to identify the signature of ICM close to $\mu=0$.
\acknowledgments{}
 We acknowledge A. A. Zyuzin for helpful discussions. S.F.I. acknowledges the financial support from {\color{magenta} Academy of Finland} Grant No. 308339. P.D. acknowledges financial support from the {\color{magenta} Knut and Alice Wallenberg Foundation} through the {\color{magenta} Wallenberg Academy Fellows program} awarded to Annica M. Black-Schaffer.  
  
\bibliography{bibfile}{} 
\end{document}